# A Unified Interactive Framework for Real-Time Three-Dimensional Cellular Automata Simulation in XR

Ashkan Rezaei[1,2], Azam Bastanfard[1,2*], Mohammad Erfan Safabakhsh[1,2]

1. *Department of Computer Engineering, Ka.C., Islamic Azad University, Karaj, Iran*

2. *The Research Center for Interactive Technologies and Extended Reality, Karaj Branch, Islamic Azad University, Karaj, Iran*

* Corresponding author: Azam Bastanfard

**Abstract**

Integrating three-dimensional cellular automata (3D CA) into extended reality (XR) requires coordinated simulation, volumetric data management, rendering, and interaction under variable hardware constraints. This paper presents a hybrid study combining a critical review of existing approaches with a unified architecture for real-time 3D CA in XR. The proposed five-layer architecture integrates voxel and rule management, parallel computation, interoperability middleware, visualization, and XR interaction. Its key principles include hardware-aware execution, asynchronous simulation-rendering, adaptive fidelity, and cross-engine interoperability. An analytical evaluation framework is defined to assess performance, latency, scalability, memory efficiency, simulation correctness, and interaction quality. The proposed framework provides a foundation for developing portable and interactive 3D CA systems across desktop and mobile XR platforms.

Email addresses: ashkan.rezaei0702@iau.ir (Ashkan Rezaei),

bastanfard@iau.ac.ir (Azam Bastanfard ),

mohammaderfan.safabakhsh@iau.ir (Erfan Safabakhash),

## Introduction

Cellular automata (CA) have evolved from foundational models of local, rule-governed systems into practical tools for investigating complex spatiotemporal processes across scientific and engineering domains. Their appeal comes from a simple computational structure: a discrete space is populated by cells whose states evolve through local transition rules. Despite this simplicity, collective interactions can generate patterns and dynamics that are difficult to derive analytically. Recent work has explored CA in three-dimensional visualization and immersive exploration, as well as in domains such as environmental modelling, fluid-like behaviour, biological processes, and molecular-scale interactions [3, 7, 16, 33, 39, 49].

Historically, the computational cost of updating large cellular grids restricted the size and temporal resolution of simulations. General-purpose GPU computing and other forms of parallel execution have changed this landscape by mapping cell updates to parallel hardware. GPU-based CA approaches have been studied for hexagonal and three-dimensional automata, medical volumetric data, indoor particulate-matter modelling, urban expansion, and large-scale microstructure simulation [9, 11, 12, 36, 42, 43]. More recent work investigates tensor-core execution and bit-plane or bitwise representations, illustrating that performance depends not only on the number of processing units but also on the neighbourhood structure, state representation, memory layout, and transition function [4, 21]. These developments provide important computational building blocks, but they do not by themselves guarantee an interactive XR experience.

The distinction is important because an XR application must satisfy several coupled requirements. A simulation can update cells rapidly yet produce an unstable display because data transfer, synchronization, or rendering becomes the bottleneck. Conversely, a visually smooth scene may display stale simulation states if the update process is decoupled without an explicit consistency policy. Immersive systems also introduce constraints that are less central to conventional offline simulation: viewpoint-dependent rendering, motion-to-photon latency, continuous head and controller tracking, limited thermal and power budgets on standalone devices, and the need for interaction techniques that make three-dimensional state changes understandable. Research on scientific visualization in VR and large-scale data visualization demonstrates the value of immersive exploration, while also showing that integration between specialized visualization software and VR environments is a distinct engineering problem [10, 22, 25, 29].

A further gap concerns interoperability. Scientific CA solvers and XR engines are often developed around different data models, execution assumptions, and programming interfaces. A simulation may be implemented in a scientific computing environment, while the immersive front end may be built in a game engine or browser-based graphics platform. Frameworks such as MINT illustrate how visualization systems and VR environments can be loosely coupled without porting all visualization algorithms into the XR engine; WebGPU-oriented engines and WebXR toolkits offer complementary routes to cross-platform rendering and interaction [10, 17, 31]. Nevertheless, a portable CA-to-XR pipeline requires an explicit contract for grid dimensions, state formats, coordinate systems, time steps, buffer ownership, synchronization, and rendering semantics. Without such a contract, engine-specific adaptations can undermine reproducibility and increase maintenance effort [2, 8, 13].

The interaction problem is equally significant. A 3D CA is not merely a volume to be viewed; it is a dynamic state space that users may need to inspect, slice, annotate, perturb, compare, or manipulate. Existing work has explored VR-based exploration of 3D CA, head-mounted visualization of automata and tilings, immersive scientific-image exploration, and interactive agent-based simulation [3, 6, 29, 34]. These studies suggest opportunities for spatial interaction, but also raise questions about navigation efficiency, cognitive load, and whether interaction patterns familiar from desktop applications transfer effectively to immersive settings.

This paper addresses the intersection of these issues. It does not claim that one existing system already solves the entire problem. Instead, it synthesizes relevant work and proposes a unified architecture intended to coordinate parallel CA execution, volumetric data handling, cross-engine exchange, rendering, and immersive interaction. The contribution is therefore architectural and methodological: the paper defines the layers, interfaces, design principles, and evaluation criteria needed to build and assess a real-time 3D CA environment in XR.

## Literature Review

Prior work establishes several complementary foundations for the proposed research. Foundational visualization research examined multiscale CA data and GPU-based visual simulation, while subsequent work extended GPU computation to hexagonal and three-dimensional automata [14, 15, 19, 11, 12]. Application-oriented studies have used CA for medical volumetric processing, indoor particulate-matter modelling, urban expansion, traffic-flow visualization, fluid-like interaction, environmental dispersion, and fire-spread simulation [9, 16, 37, 38, 40, 41, 42, 43].

Recent computational work explores tensor-core execution, bit-plane encoding, vectorization, and alternative parallel architectures, indicating that the most suitable acceleration strategy depends on neighbourhood size, state encoding, memory traffic, and rule structure rather than on GPU use alone [4, 7, 21, 36]. In parallel, immersive visualization research has moved from direct exploration of three-dimensional CA and tilings toward broader scientific visualization and large-scale data exploration in VR, while interactive systems have investigated spatial manipulation, collaborative exploration, and lightweight WebXR interaction [3, 10, 17, 22, 25, 29, 34]. Cross-platform rendering and interoperability work, including WebGPU-based engines and middleware-oriented integration, provides mechanisms for separating scientific computation from immersive presentation, while recent XR platforms demonstrate the feasibility of interaction on a wider range of devices [2, 8, 13, 31, 39]. Finally, neural and adaptive CA research suggests a broader transition from fixed hand-designed rules toward learned or dynamically modified spatiotemporal behaviour [27, 28, 32]. Taken together, these studies provide strong individual building blocks, but they are distributed across computational, visualization, interaction, and interoperability communities. The critical gap is therefore not the absence of GPU CA, VR visualization, or cross-platform rendering in isolation; it is the absence of a common, explicitly specified architecture and evaluation protocol that connects these capabilities while preserving simulation correctness and immersive responsiveness.

Studies have explored fuzzy-based data representation and computational efficiency through the design of a fuzzy emotional speech corpus and real-time speech emotion recognition using a reduced set of discriminative features [50, 51]. These studies provide relevant methodological perspectives on managing uncertainty and reducing computational complexity, which can inform the development of adaptive state representations and lightweight processing strategies in real-time cellular automata systems. More directly, research has investigated the generation of Ma'qeli patterns using block cellular automata, demonstrating the applicability of structured neighborhood configurations and rule-based state transitions for spatial pattern generation [52]. Collectively, these studies provide complementary foundations for developing modular cellular automata architectures that combine structured transition rules, flexible state representations, and computational efficiency.

From the perspective of immersive interaction and visualization, research has investigated gamified extended reality and pseudo-haptic feedback for laparoscopic skill training, highlighting the importance of precise spatial manipulation, real-time feedback, and responsive interaction mechanisms in XR environments [53]. Similarly, an interactive lighting and shadowing system for augmented reality models has demonstrated the role of user-controlled visual parameters in improving the representation and realism of three-dimensional virtual objects [54]. Although these studies address different application domains, their contributions provide relevant methodological insights for designing the interaction and visualization components of the proposed CA-XR framework, particularly in supporting real-time manipulation, responsive visual feedback, and immersive exploration of dynamic three-dimensional data.

Table 1: Summary of the reviewed studies across five performance dimensions.

| Study | Computational Throughput Efficiency | Frame Rate Stability Performance | Latency Overhead Reduction | Scalability Limit Capacity | Interaction Fluidity Assessment |
|---|---|---|---|---|---|
| (Gobron et al., 2011) | GPU processing is two hundred times faster than identical CPU algorithms | Real-time visualization of large 3D matrices is achieved through GPU acceleration | Original encoding methods transmit large CA functions to the GPU in real time | Large 3D matrices are processed efficiently using graphics processor unit hardware | Interactive user interface allows testing with different project ideas and configurations |
| (Pajouheshgar et al., 2024) | MeshNCA synthesizes dynamic textures on 3D meshes in real time | WebGL shading language implementation ensures consistent performance on personal computers | Forward pass implementation allows for immediate texture synthesis on unseen 3D meshes | Operates on non-grid structures like vertices of a 3D mesh without UV maps | Includes texture density, orientation controls, grafting brushes, and motion speed direction controls |
| (Pajouheshgar et al., 2023) | Real-time 3D texture synthesis is achieved using WebGL shading language implementation | Trained models showcase performance on personal computers and smartphones via interactive demos | Forward pass implementation enables real-time synthesis on any mesh after initial training | Generalized cellular automata operate on non-grid structures like vertices of 3D meshes | Allows user interactions including texture density, orientation control, grafting, and motion speed |
| (Arevalo et al., 2022) | Audio spatialization renders 255 sources without dropouts in virtual reality environments | VR version rated excellent in attractiveness compared to traditional desktop benchmark data | Accuracy of spatialization is based on actual distance measurements rather than approximations | Maximum number of audio sources supported is 255 by the underlying software | Subjective evaluation indicated no significant differences in user experience between VR and PC |
| (Trapeznikova et al., 2024) | Unity 3D engine enables efficient implementation and visualization of traffic flow models | Graphical interface allows users to set initial parameters and obtain traffic statistics | Cross-platform engine supports construction of lattices of arbitrary sizes in 3D space | Nagel-Schreckenberg model generalized for heterogeneous flows on multi-lane highways underlies the program | Graphical interface allows users to set initial system parameters and obtain traffic statistics |
| (Shahir & Beigi, 2026) | 3D Neural Cellular Automata synthesize intricate structures while maintaining high computational efficiency | Model enables efficient training and robust generalization across a variety of volumetric patterns | Architecture supports learning and sustaining complex structures in 3D environments in real time | 3D extension of neural cellular automata allows for scaling to volumetric phenomena | Framework integrates machine learning with mechanistic modeling to learn complex spatio-temporal dynamics |
| (Sophie, 2023) | Simulator created using an AR-type head-mounted display for 3D cellular automata visualization | Visualized 3D version of Penrose tiling using head-mounted display technology | Real-time simulation of various 3D CAs including life games in AR environments | Supports simulation of various 3D cellular automata and complex quasiperiodic tiling structures | Head-mounted display provides immersive visualization of 3D cellular automata and tiling models |
| (Gobron et al., 2008) | GPU-accelerated method computes CA close to a thousand times faster than CPU | Real-time rendering of cellular automata applied to hexagonal grids on graphics hardware | Novel encoding transmits large CA key-codes to the graphics card in real time | Efficiently identifies emerging behaviors even for large data sets on different hardware | Flexible model allows for easy identification of emerging behaviors in large data sets |
| (Yu & Chang, 2023) | GPU-based cell parallelization enhances efficiency by up to 7695 percent over CPU | Maintains simulation accuracy while significantly enhancing efficiency for indoor concentration modeling | Parallelization of transport mechanisms performed on GPUs instead of sequentially on CPU | Simulations performed on i9 PC with Intel UHD Graphics 630 graphic card | Parallelized CA approach proves useful for real-time 3D indoor distribution modeling |
| (Brabec et al., 2026) | Bit-plane data encoding enables efficient bitwise vectorization and temporal blocking optimizations | Offers speedup of two orders of magnitude compared to baseline CUDA implementations | Cellato abstraction allows simple CA rule definitions using C++ templates for users | Versatile approach implemented for 10 different CA models to demonstrate performance | Maintains code simplicity and ease of use for the end users during simulation |
| (Navarro et al., 2024) | Tensor core approach exhibits constant time throughout the entire neighborhood radius range | Outperforms all other approaches starting from radius three for cellular automata simulations | GPU tensor core approach accelerates CA with weighted summation of neighborhood cells | Reaches speedups of up to 101 times over a GPU baseline implementation | Most energy efficient approach for neighborhood radius greater than or equal to five |
| (Zhang, 2026) | Graph structure facilitates efficient processing and analysis of complex cellular automata systems | Dynamic rule adaptation mechanism continuously updates parameters based on observed simulation dynamics | Graph-based representation addresses limitations of computational complexity in traditional CA simulations | Effectively handles complex rule sets and large spatial scales in CA systems | System responds to evolving patterns to achieve greater accuracy in simulating behaviors |
| (Pandit & Kudithipudi, 2026) | Mini-CA kernels enhance parallel throughput and enable input scalability for visual streams | Achieves state-of-the-art accuracy while significantly reducing computational latency compared to baselines | Decomposes monolithic CA systems into specialized, computationally efficient units for processing | Extends CA applicability beyond binary inputs to process spatio-temporal sequences and images | Integrates directly into CNN architectures replacing standard operations with dynamic rule-based processing |
| (Zhang, 2026) | Dynamically adjusted rules allow system to maintain compact and accurate geometric representation | Adaptive mechanism allows system to react to changes in real-time geometric problems | CA system where each cell represents a geometric element for efficient processing | Efficiently processes evolving geometric datasets in dynamic environments where elements undergo transformations | Offers a flexible framework for diverse geometric applications including robotics and computer vision |

*Continued on next page*

Table 1 – *continued from previous page*

| Study | Computational Throughput Efficiency | Frame Rate Stability Performance | Latency Overhead Reduction | Scalability Limit Capacity | Interaction Fluidity Assessment |
|---|---|---|---|---|---|
| (Svyetlichnyy & Banaś, 2026) | Linked-list acceleration achieves highest performance among considered frontal cellular automata variants | Memory-access conflicts eliminated by separating information transfer into directional flows | Parallel CPU and GPU execution reduces computational time for large-scale simulations | Demonstrates good scalability for CPU and GPU implementations in microstructure evolution | Enables safe parallel processing of multiple linked lists while preserving simulation correctness |
| (Xu et al., 2026) | Voxel-native GPU-parallel architecture completes simulations of millions of voxels within minutes | Stable transport behavior and encouraging agreement with CFD benchmark simulation results | Interactive runtime performance suitable for building management and environmental monitoring applications | Supports urban-scale 3D city models containing up to 13.9 million active voxels | Provides efficient voxel-based approximation of gas dispersion suitable for interactive environmental analysis |
| (Heintz et al., 2017) | High framerate of 100 FPS achieved for CA running on the CPU | Parallelization and optimization allow for high framerate in fluid simulation applications | Short computation time makes the simulation interesting for games in virtual reality | Supports 1763 cells due to parallelization and optimization techniques in the simulation | Realtime fluid simulation allows for interaction with different kinds of fluids simultaneously |
| (Harris et al., 2002) | Programmable graphics hardware used for real-time visual simulation of diverse dynamic phenomena | Interactive framework built for experimenting with CML simulations running on graphics hardware | Pixel-level programming implements next-state computations on lattice nodes and their neighbors | Stores lattice values in a texture to produce interactive visual simulations of phenomena | Integrated into interactive 3D graphics applications for building and experimenting with simulations |
| (Harris et al., 2005) | Implementation requires nine passes per slice for the 3D boiling simulation iteration | Runs at 45 frames per second in a virtual environment built on game engine | Trade-off between optimal performance and flexibility in the CMLlab framework design | Supports a range of resolutions for 2D and 3D boiling and convection simulations | Interactive framework allows constructing simulations from a set of general purpose operations |
| (Davison et al., 2018) | Sketch-based discrete element texture synthesis used to paint large agent arrangements | Enables new levels of interactivity for design, simulation, and exploration of agents | Allows users to pause the simulation and modify agents with sketch-based tools | Supports large number of agents in 3D agent-based simulations for interactive painting | Sketch-based generative brushes allow reconfiguring agents to match examples created in palette |
| (Ohno & KAGEYAMA, 2024) | Allows for interactive analysis of 3D scalar and vector fields in VR | Enlarge and analyze specific regions of large-scale simulation data in real time | Modern HMDs provide a cost-effective alternative to room-sized CAVE display systems | Designed for large-scale simulation data visualization in virtual reality environments | Interactive analysis of 3D fields enabled through Unity and OpenXR plugin integration |
| (Geringer et al., 2024) | Image-based approach integrates scientific visualization into VR without explicit porting | Enables enriching VR-based scientific data exploration with novel immersion and interaction techniques | Separation of concerns enables researchers to independently work with existing software frameworks | Facilitates integration of visualization algorithms from software frameworks directly into VR | Enables collaborative VR environment implementation for scientific data exploration and visualization |
| (Kushwah et al., 2026) | Real-time rendering and automated feedback mechanisms enhance student engagement and practical understanding | Measurable improvements in visualization accuracy and interaction efficiency compared to traditional approaches | Cloud-based accessibility and drag-and-drop interaction strengthen conceptual understanding of engineering concepts | Modular architecture supports future expansion into AI-driven adaptive learning and collaborative environments | Enables learners to interact with virtual components and execute simulations through immersive environments |
| (Sung et al., 2025) | GPU-based parallel processing framework enhances real-time performance through sophisticated collision detection | Achieves stable real-time frame rates for complex models like Stanford Bunny and Armadillo | Storing collision data in GPU-accessible 2D textures improves FPS by 117 percent | Robust, scalable framework for real-time physics simulations in extended reality applications | Employs AABB-based bounding volume hierarchy for precise triangle-level interactions in XR |
| (Waseem & Hong, 2025) | Achieves throughput rates up to 168,600 particles per millisecond in fluid simulations | Maintains consistent 5.7 to 6.0 ms frame times across varying particle counts | Novel fusion of Count Sort with Parallel Prefix Scan optimizes neighborhood search | Supports real-time fluid simulations with up to 1 million particles in Unity | Maintains interactive frame rates above 30 FPS for large-scale fluid simulations |
| (Olivier & Oded, 2022) | Dedicated cellular automata implemented on reconfigurable hardware for high-performance simulation | Bypasses bottleneck by sending only a small subset of state into the simulation | Torchlight concept allows simulation on a micro level outside the locus of visibility | Enables larger and more fine-grained simulation for training neural networks for games | Interfaces with a 3D game engine to translate raw simulator data to graphics |
| (Ahmed, n.d.) | VCELL framework allows different simulation models to receive real-time data to interact | Improves models' design and reduces scenario development, modification, and validation time | Integrates 3D visualization, sensor data, and modeling and simulation in real time | Supports large-scale programs for emergency and disaster simulation in time and space | Allows cellular agent model and collaborative 3D visual agent model to interact |
| (Zhang, 2026) | WebGL-based framework renders dense array of over 4,000 dual-encoded physical glyphs | Significant negative correlation between 6-DOF interaction fluidity and task frustration | Pose-volume dual-mapping mechanism represents field direction and magnitude in real time | High-performance simulation framework supports dense array of physical glyphs for education | Non-STEM participants achieved 62.2 percent accuracy in complex tensor topology diagnostics |
| (Rodríguez et al., 2025) | Engine ensures consistent rendering fidelity across desktop and web environments | Full support for OpenXR and experimental WebXR-WebGPU bindings for immersive applications | WebAssembly compilation and rendering backend adaptation via Google's Dawn framework | Cross-platform graphics engine supports WebGPU across desktop and web environments | Flexible foundation for future research, education, and experimental development in immersive graphics |

*Continued on next page*

Table 1 – *continued from previous page*

| Study | Computational Throughput Efficiency | Frame Rate Stability Performance | Latency Overhead Reduction | Scalability Limit Capacity | Interaction Fluidity Assessment |
|---|---|---|---|---|---|
| (Faldor & Cully, 2024) | Speeds up simulations up to 2,000 times faster than traditional methods | Modular design through a user-friendly interface supports discrete and continuous automata | High-performance open-source library designed to accelerate cellular automata research | Supports any number of dimensions for cellular automata research and applications | Offers cutting-edge performance and flexibility through a wide range of benchmarks |
| (Zhou et al., 2026) | Achieves 8.15 times acceleration compared to traditional OpenFOAM CFD software | Tight integration of CFD model, game engine, and GIS data ensures efficiency | Unity compute shaders achieve fast, fully parallelized computation while maintaining physical fidelity | Expanded spatial extent by incorporating data from multiple GIS sources for urban contexts | Ensures user interactivity, scene modifiability, and intuitive visualization for urban planning |
| (Pandolfi et al., 2025) | Immersive software accelerates visualization and exploration of 3D dense arrays and meshes | Multi-user capabilities allow multiple users to share data and visualize together | Integration with HPC resources grants remote processing of large-scale data streamed into VR | Dynamically load and manipulate scientific records to dive into the structure of data | Unique integration combining VR, HPC, and AI-driven object modeling for scientific visualization |
| (Czapczyński & Dziedzic, 2026) | Multithreaded rendering with multipass rendering achieved highest frame rate in VR | Best overall metrics observed for configuration combining multithreaded and multipass rendering | Use of multiview rendering led to reduction in average CPU utilization | Tested four versions of two Unity applications using Meta Quest Pro headset | Performance metrics compared across twelve valid build configurations for VR applications |
| (Hadjar et al., 2026) | Achieves fast loading times and stable rendering performance across different conditions | Five-layer framework ensures real-time synchronization between physical and virtual systems | Implemented using Three.js and WebXR for browser-based performance without specialized software | Supports domain-agnostic applications and optimized 3D visualization for digital twins | Enables users to influence system behavior through AR/VR interfaces and intelligent analytics |
| (Shen et al., 2026) | Interactive web-based platform for real-time demonstration of cardiac reentry dynamics | Students manipulate key model parameters and immediately observe how these shape wave dynamics | Network is built from student devices each representing a model cell coupled to neighbors | Simplified cellular automaton model used for teaching excitable-media dynamics | High engagement reported in pilot classroom deployments with undergraduate life-science students |
| (Park & Choi, 2025) | Unity supports real-time simulations for biological and medical research visualization | Blender excels in making highly accurate static models for biological processes | Unreal Engine provides stunning visuals for detailed 3D biological models | Comparative analysis of three popular tools for creating detailed 3D visuals | Explored how three tools help scientists create detailed 3D visuals of cells |
| (Zhang, 2026) | Minimizes communication overhead and maximizes computational efficiency in distributed environments | Novel synchronization strategy based on Byzantine Fault Tolerance guarantees accuracy and robustness | Employs CA to model system dynamics at a local, granular level | Scalable simulation of complex systems including urban dynamics and traffic flow | Ensures consistency among distributed simulation nodes using consensus algorithms |
| (Ganji, 2026) | Sustains true immersive experience under tight latency budgets for XR development | Keeps inter-module synchronization below 1 ms in responsive driving simulator | Execution optimization for true immersive experience through module abstraction and orchestration | Supports real-time volumetric video conferencing while reducing bandwidth by 98 percent | Resilient to mid-session module replacement and flexible for resource efficiency |
| (Xu et al., 2026) | Voxel-native parallel memory layout enables constant-time neighbor access and efficient updates | Predictable runtime scaling and practical performance for domains exceeding one million voxels | Physics-informed, GPU-accelerated framework for simulating 3D fire propagation in WUI environments | Evaluated using voxelized model of Liverpool and tested across high-performance computing systems | Enables physically grounded simulation without relying solely on empirical or probabilistic rules |
| (Zeng et al., 2024) | Execution time reduced by more than 98 percent compared to conventional CA models | Convolution function in DL library used to calculate neighborhood effect efficiently | GPU-convolution-CA model leverages similarity between neighborhood effect and convolutional process | High-performance CA model constructed for urban expansion simulation based on GPU | Efficient application programming interfaces handle tasks of interest for CA simulation |
| (Herruzo et al., 2026) | Median latency of 470 ms and 95th-percentile latency under 812 ms | Interactive frame rates at production workloads for mobile performance constraints | End-to-end cyber-physical architecture supports continuous sensing ingestion and spatial inference | Spatializes localized environmental sensing at city scale using 260 weather stations | Walk-up access via QR or URL on commodity mobile devices without installation |
| (Gao et al., 2011) | Segmentation accelerated by GPU for real-time response in surgery simulation | Evolution can be observed and guided with volume rendering results in VR | User interactive segmentation method requires few interactions for high precision | Supervised by pyramids according to selected resolution for medical volumetric data | Labels image generated in higher level Gaussian pyramid extended to lower levels |
| (Hentschel et al., 2026) | WebXRIT consistently outperforms default XR Interaction Toolkit in frame rates | Achieves significantly higher frame rates in multiple configurations up to 78 FPS | Lightweight framework minimizes runtime overhead for WebXR-based training applications | Integrates with open-source WebXR Export plugin and follows component-based interaction architecture | Enables performant and accessible IVR training applications on the web |
| (Zhang, 2026) | Novel parallel algorithm significantly enhances simulation speed and scalability | Distributes simulation workload across multiple processors for faster exploration of parameter spaces | Rapidly tests and evaluates cellular automata simulations in distributed computing system | Overcomes limitations of computational complexity in scaling to larger parameter spaces | Demonstrates significant improvement in simulation speed compared to traditional methods |

*Continued on next page*

## Critical Analysis and Synthesis

The current body of literature demonstrates a robust shift toward GPU-accelerated architectures and cross-engine interoperability to overcome the computational bottlenecks inherent in real-time 3D cellular automata (CA) simulations. While recent frameworks achieve significant performance gains through parallelization and novel memory layouts, there remains a notable tension between maintaining high-fidelity volumetric simulation and ensuring the low-latency responsiveness required for immersive extended reality (XR) environments.

| | Strengths | Weaknesses / Limitations |
|---|---|---|
| **Computational Efficiency** | GPU cell parallelization, tensor cores and bit-plane encoding give speedups of orders of magnitude over CPU implementations. *(Gobron et al., 2011; Yu & Chang, 2023; Brabec et al., 2026; Navarro et al., 2024; Zeng et al., 2024)* | Many solutions are domain-specific or tied to proprietary hardware, hindering a universal, scalable architecture. *(Svyetlichnyy & Banas, 2026; Ganji, 2026; OpenCAL++, 2023)* |
| **Scalability** | Voxel-native architectures and linked-list management support millions of active cells at interactive frame rates on commodity hardware. *(Xu et al., 2026; Waseem & Hong, 2025)* | Urban/large-scale datasets force fidelity trade-offs or "locus of visibility" techniques that limit global scope. *(Olivier & Oded, 2022; Herruzo et al., 2026; Hnat & Chapman, 1999)* |
| **User Interaction** | Sketch-based generative brushes, multi-user collaboration and spatialized audio enhance exploration of complex dynamic systems. *(Arevalo et al., 2022; Davison et al., 2018; Pandolfi et al., 2025; Wonnacott et al., 2026)* | Users struggle to explore the full extent of 3D simulation universes in VR compared with desktop; navigation paradigms need work. *(Arevalo et al., 2022)* |
| **Cross-Engine Interoperability** | Open standards (WebXR) and middleware (mint) enable integration into diverse game engines without extensive porting. *(Geringer et al., 2024; Rodriguez et al., 2025; Hadjar et al., 2026)* | Interoperability overhead can conflict with strict XR latency; simulation-rendering synchronization needs careful optimization. *(Ahmed, n.d.; Ganji, 2026)* |
| **Methodological Rigor** | Standardized benchmarking (frame rate, latency, throughput) and validation against established CFD solvers. *(Yu & Chang, 2023; Xu et al., 2026; Sung et al., 2025; Hentschel et al., 2026)* | No standardized benchmarking metrics tailored to 3D CA in XR, so cross-architecture comparison is difficult. *(Brabec et al., 2026; Ortiz et al., 2022)* |

Figure 2: Unified framework for scalable, interactive 3D CA systems must prioritize the integration of hardware-agnostic parallelization strategies with standardized interaction toolkits to bridge the gap between scientific simulation and user-centric exploration.

Overall consensus: GPU-accelerated architectures are essential for real-time 3D cellular automata (CA) and volumetric simulation, with substantial speedups over CPU-based implementations. (Gobron et al., 2011; Gobron et al., 2008; Yu & Chang, 2023; Navarro et al., 2024; Xu et al., 2026; Sung et al., 2025; Zeng et al., 2024)
Overall divergence: studies differ on the optimal software abstraction and hardware deployment, from specialized low-level kernels to high-level game-engine integration, reflecting trade-offs among simulation fidelity, cross-platform interoperability and target hardware (mobile XR vs. high-performance desktop).

| Comparison criterion | Studies in agreement | Studies in divergence | Potential explanations |
|---|---|---|---|
| **Computational Efficiency** | GPU-based parallelization significantly outperforms CPU execution. *(Gobron et al., 2011; Gobron et al., 2008; Yu & Chang, 2023; Sung et al., 2025; Zeng et al., 2024)* | Approaches vary from custom CUDA/OpenCL kernels to high-level engine-based compute shaders. *(Yu & Chang, 2023; Xu et al., 2026; Zhou et al., 2026; Zeng et al., 2024)* | Differences in hardware targets (commodity vs. HPC) and in the complexity of the underlying CA rules influence the choice of implementation. |
| **Scalability Strategies** | Large-scale simulations benefit from spatial partitioning or “locus of visibility” techniques. *(Svyetlichnyy & Banaś, 2026; Ohno & KAGEYAMA, 2024; Olivier & Oded, 2022; Xu et al., 2026)* | Some frameworks prioritize monolithic grid updates, whereas others use graph-based or agent-based decomposition. *(Zhang, 2026; Pandit & Kudithipudi, 2026; Svyetlichnyy & Banaś, 2026)* | Discrepancies arise from the application domain, such as urban-scale fire spread versus localized molecular dynamics. |
| **Interaction Paradigms** | Immersive XR environments enhance the exploration of complex 3D data through spatial interaction. *(Arevalo et al., 2022; Ohno & KAGEYAMA, 2024; Pandolfi et al., 2025; Wonnacott et al., 2026)* | Interaction methods range from direct manipulation and painting to parameter-driven steering and collaborative multi-user sessions. *(Davison et al., 2018; Shen et al., 2026; Gao et al., 2011; Wonnacott et al., 2026)* | Variations in user goals, from educational demonstration to scientific analysis, dictate the required level of interaction precision. |
| **Cross-Engine Interoperability** | Integration with game engines such as Unity is a preferred path to interactive visualization. *(Trapeznikova et al., 2024; Ohno & KAGEYAMA, 2024; Kushwah et al., 2026; Zhou et al., 2026; “3D Visualization of Forest Fire Spread M…”, 2023)* | Some systems prioritize web-based, zero-install architectures (WebGL/WebGPU) over native engine plugins. *(Pajouheshgar et al., 2024; Rodríguez et al., 2025; Hadjar et al., 2026; Herruzo et al., 2026)* | Divergence reflects the priority placed on accessibility and platform-agnostic deployment versus the high-fidelity requirements of native VR applications. |

**Figure 3.** Agreement and divergence across the reviewed studies on 3D cellular automata in extended reality. For each comparison criterion, the figure summarizes the points of consensus, the points of divergence (with supporting references), and the potential explanations for the observed differences.

## METHODOLOGY

The framework is designed around the premise that a real-time CA-XR system is a pipeline, not a single simulation kernel. Its objectives are to (i) preserve the semantics of the CA transition model, (ii) make computation portable across hardware classes, (iii) minimize unnecessary data copies and synchronization, (iv) keep rendering responsive when simulation and display rates differ, (v) allow different XR engines to consume the same simulation stream, and (vi) expose meaningful tools for volumetric exploration.

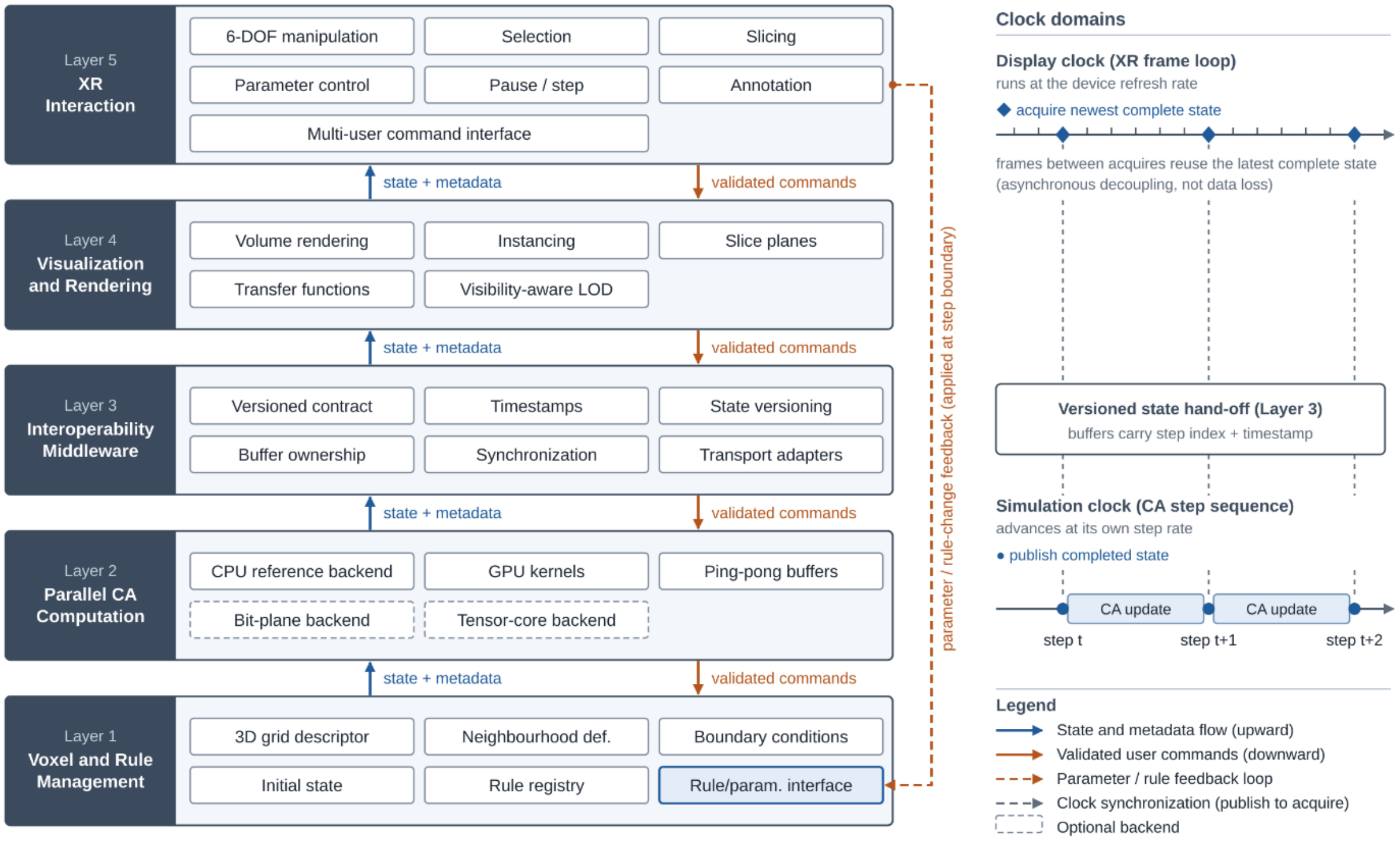

Figure 4. Unified CA-XR architecture

## Five-layer architecture

Layer 1 — Voxel and Rule Management. This layer owns the spatial grid, cell-state schema, neighbourhood definition, boundary conditions, transition-rule identifier, initial conditions, and simulation parameters. It supports dense grids for regularly occupied volumes and sparse or chunked representations when the active region is limited. A grid descriptor contains dimensions, physical cell spacing, origin, axis convention, state type, and boundary policy. The rule interface is independent of the compute backend so that a rule can be executed on CPU or GPU without changing its scientific definition.

Layer 2 — Parallel CA Computation. This layer executes the transition function. For a synchronous automaton, the state at step t + 1 is computed from the state at step t using separate input and output buffers. This ping-pong arrangement avoids reading values that have already been updated in the same step. Backend implementations may include CPU multithreading, conventional GPU kernels, bit-plane or bitwise kernels, tensor-core approaches where the rule structure permits, or other available parallel mechanisms. The framework should not force tensor cores or a particular API; backend capabilities are queried and benchmarked before selection [4, 7, 21, 36].

Layer 3 — Interoperability Middleware. This layer separates the scientific solver from the XR engine. It manages the exchange contract, resource ownership, state versioning, timestamps, buffer availability, and synchronization. Possible transport mechanisms include shared memory, GPU-compatible shared resources, inter-process communication, or a network transport when clients are distributed. The protocol should be transport-neutral: the semantic message remains the same even when its physical transport changes. This principle is consistent with middleware approaches that connect scientific visualization systems and VR environments without requiring all algorithms to be reimplemented inside one application [10].

Layer 4 — Visualization and Rendering. This layer converts simulation state into visual representations. It may use instanced geometry, volume rendering, slice planes, surface extraction, transfer functions, or hybrid representations. The rendering path should avoid unnecessary CPU readback when a compatible GPU resource can be shared. It should also support visibility-aware prioritization and visual LOD. Importantly, rendering LOD changes the visual representation, not the simulation state, unless a separately declared approximate simulation policy is enabled.

Layer 5 — XR Interaction. This layer receives headset, controller, hand, gaze, or desktop input and translates it into high-level commands. Commands include selecting a region, moving a slice plane, changing a visual transfer function, pausing or stepping the simulation, modifying parameters, placing annotations, and applying a controlled intervention. Each intervention is timestamped and validated by the simulation layer before being committed. The interaction layer also supports navigation and inspection tools such as 6-DOF manipulation, focus-and-context views, and region-of-interest controls.

Conceptual data flow

The normal execution cycle follows this sequence:

1. The rule manager validates the grid descriptor, neighbourhood, boundary conditions, and parameter set.
2. The scheduler chooses a compute backend and allocates the required state buffers.
3. The CA engine computes one or more state transitions into an output buffer.
4. The middleware publishes a versioned state descriptor and makes the associated buffer available to the renderer.
5. The visualization layer updates the visible representation using the newest complete state that meets its freshness policy.
6. The XR layer renders the scene and records display and interaction timestamps.
7. User commands are validated, queued, and applied at a defined simulation boundary to preserve deterministic update semantics.

This cycle separates the simulation clock from the display clock. The simulation may advance at a fixed step, while the XR display runs at the device refresh rate. The renderer can display the latest completed state without forcing the simulation to complete a new update during every display frame. The interface must make the displayed state time visible when the difference between simulation time and wall-clock time becomes meaningful.

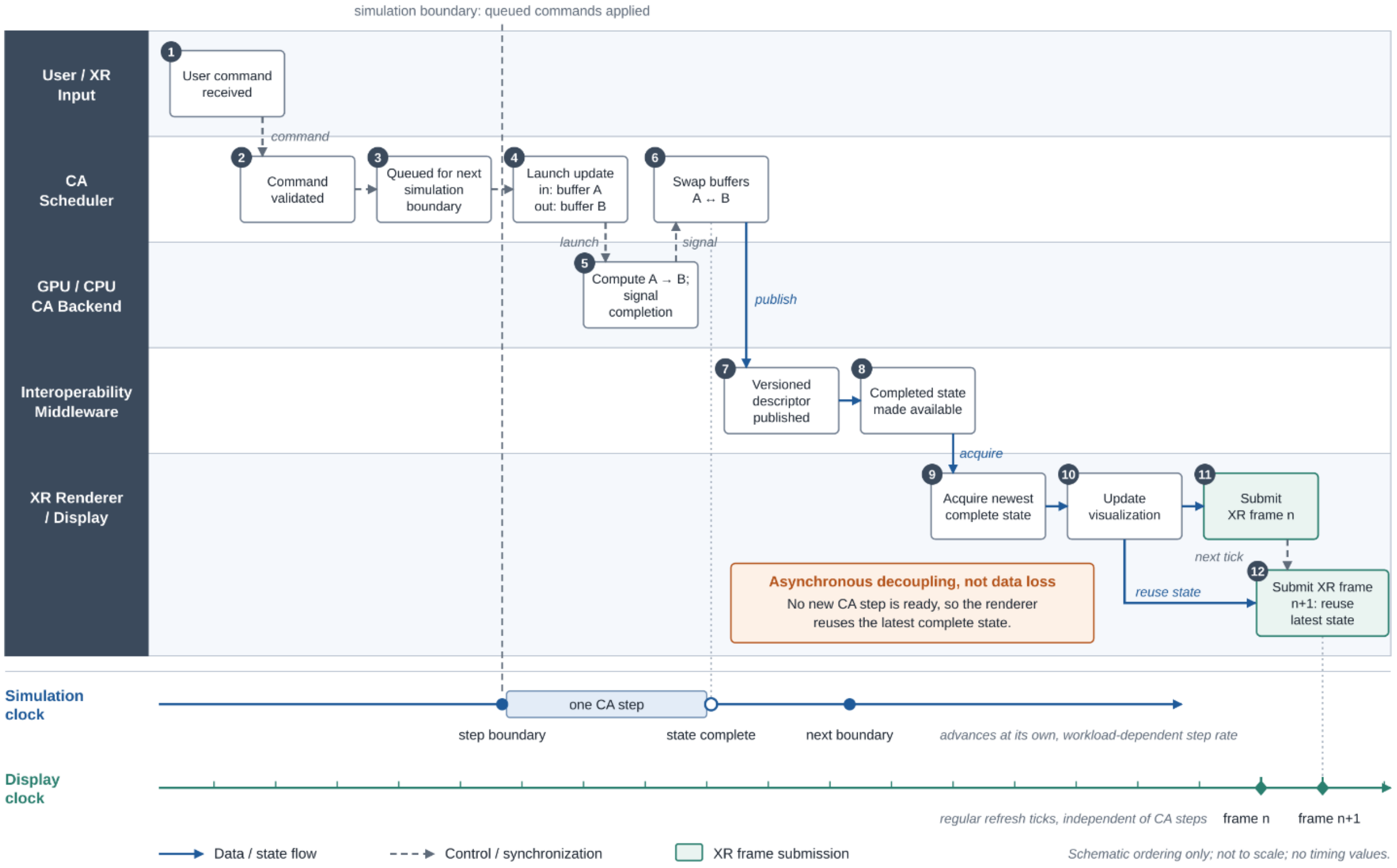

*Figure 5. Asynchronous simulation-to-display data flow*

## Formal CA model

Let the three-dimensional lattice be

$$\Lambda = \{(x,y,z) \in \mathbb{Z}^3 : 0 \le x < N_x, 0 \le y < N_y, 0 \le z < N_z\} \quad (1)$$

Each cell $c \in \Lambda$ has a state at discrete simulation step t, where S is the state space and t denotes the discrete simulation step:

$$s_c^t \in S,\ c \in \Lambda,\ t \in \mathbb{N}_0 \quad (2)$$

Let $\mathcal{N}(c)$ denote the neighbourhood of cell c, and let $\theta$ contain rule parameters. A synchronous CA update is represented as

$$s_c^{t+1} = f(s_c^t, \{s_n^t : n \in \mathcal{N}(c)\}; \theta) \quad (3)$$

The function f is independent of the selected execution backend. For a standard synchronous model, all output cells at step t + 1 are computed from the same input state at step t. Boundary conditions are handled by an explicit policy, such as fixed, periodic, reflective, or domain-specific boundaries. The framework must record the selected policy as part of the grid and rule metadata.

The computation cost per step depends on the number of updated cells, neighbourhood size, state representation, and transition function. A simple update-work measure is

$$W_t = |A_t| \cdot |\mathcal{N}| \cdot \omega_f \tag{4}$$

where A_t ⊆ Λ is the set of cells updated at step t, $|\mathcal{N}|$ is the neighbourhood size, and ω_f is a workload descriptor for the transition rule. This descriptor is not necessarily a universal operation count; it is a way to distinguish benchmarks with different rule complexity. For sparse simulations, |A_t| may be less than the total number of cells, but the active-region policy must preserve the intended model semantics.

At initialization, the scheduler collects a capability profile: available compute backends, supported memory-sharing options, device memory budget, parallel execution limits, and current resource availability. It then selects a compatible backend and data representation. Backend selection should consider total pipeline cost, not kernel time alone. A kernel that is faster in isolation may be slower end-to-end if its output requires expensive readback or conversion before rendering.

The initial implementation should support a clear reference backend and at least one GPU backend. Optional optimized kernels—such as bit-plane encoding or tensor-core execution—should be enabled only when the transition function can be expressed correctly in the corresponding representation. The selected kernel must pass equivalence tests against the reference implementation for the same initial states, boundary conditions, and number of steps. This requirement prevents an optimization from silently changing the CA rule.

For large volumes, the grid may be divided into spatial chunks. Chunking can reduce allocation pressure and allow visibility-aware transfer, but it introduces boundary exchange and scheduling overhead. Therefore, chunk size should be treated as a benchmark parameter rather than a fixed design constant. A double- or triple-buffered state exchange can prevent the rendering thread from waiting for a buffer that is being updated. Each buffer should carry a monotonically increasing version number and simulation timestamp.

### Middleware contract and interoperability

The middleware should define a versioned data contract containing at least the fields in Table 1.

**Table 1. Versioned CA-XR data contract.**

| Field | Meaning | Validation requirement |
|---|---|---|
| schema_version | Contract version | Unknown versions are rejected or explicitly negotiated |
| grid_dimensions | Grid size along each axis (Nx, Ny, Nz) | Positive integer dimensions |
| cell_spacing | Physical spacing along each axis | Units and coordinate convention specified |

| Field | Meaning | Validation requirement |
| --- | --- | --- |
| grid_origin | Spatial origin | Coordinate system declared |
| state_type | Integer, Boolean, categorical, or numeric state | Must match the consumer representation |
| neighborhood_id | Neighbourhood definition | Identifies offsets and radius |
| rule_id and rule_version | Transition function identity | Enables reproducibility and compatibility checks |
| boundary_policy | Boundary condition | Must be explicit |
| step_index | Discrete simulation step | Monotonically increasing within a run |
| simulation_timestamp | Time associated with the state | Clock domain and units specified |
| buffer_handle or payload | State storage or serialized data | Ownership and lifetime defined |
| state_hash (optional) | Integrity or equivalence check | Computed consistently when enabled |

The contract should distinguish metadata from the state payload. A consumer should be able to determine what a buffer means without inferring its dimensions or units from engine-specific conventions. When zero-copy sharing is available, the middleware transfers a resource handle and synchronization primitive rather than copying the entire volume. When it is unavailable, the transport falls back to a documented serialization or copy path. The benchmark must measure the overhead of each path separately.

The rendering layer should prioritize the user's visible region while retaining access to the complete state. A region-of-interest policy can allocate more rendering detail to cells near the current focus, while distant cells use a lower-cost visual representation. The policy must be reversible and must not discard simulation state. If compute adaptation is introduced, the framework must label it as approximate and assess its effect on the automaton's behaviour.

The interaction layer should use commands with explicit semantics rather than direct modification of internal GPU buffers. For example, a user may select a region and request a parameter change. The command is placed in a queue and applied at a defined simulation boundary. This makes the intervention traceable and helps reproduce a session. Continuous visual controls, such as opacity or colour mapping, can be applied in the rendering layer without modifying the CA state.

For collaborative use, one authoritative simulation instance or a clearly defined distributed consistency model should own the canonical step sequence. Clients exchange interaction commands and state-version notifications rather than independently advancing divergent simulations without reconciliation. The protocol must define command ordering, conflict handling, late-joining clients, and recovery after packet loss. This extension is compatible with collaborative XR and digital-twin research, but it requires dedicated validation and is not assumed to be solved by the base architecture [13, 29, 39].

## Implementation Strategy

A practical implementation should be developed in incremental stages:

Stage A :Reference CA engine. Implement a CPU reference version with explicit rule definitions, grid metadata, boundary handling, deterministic initialization, and state serialization. This version is the correctness oracle for optimized implementations.

Stage B :GPU acceleration. Add a GPU backend using a double-buffered update model. Validate each kernel against the CPU reference and record kernel execution time separately from transfer time. Add specialized encodings only after correctness and workload suitability have been demonstrated.

Stage C :Middleware and state exchange. Implement a transport-neutral descriptor and at least one transport mechanism. Measure publication, transfer, synchronization, and buffer acquisition separately. Ensure that the renderer can detect stale, incomplete, or incompatible states.

Stage D : XR visualization. Integrate the middleware with one XR engine and implement a minimal set of visual encodings: volumetric display, slicing, selection, and state/time inspection. The rendering loop must not block on a CA step.

Stage E : Interaction and portability. Add manipulation, parameter controls, annotation, and a second engine or browser-based client to test whether the exchange contract is genuinely independent of the first engine. A single-engine implementation can validate functionality but cannot establish cross-engine interoperability.

Stage F :Benchmarking and user evaluation. Run computational benchmarks, XR frame-time tests, correctness checks, and task-based interaction studies. Report all device and configuration details.

Algorithm 1 outlines the reference execution cycle, in which user commands are applied at step boundaries, the CA state is updated through ping-pong buffers, and each completed state is published asynchronously to the renderer.

**Algorithm 1** Synchronous CA–XR execution cycle

**Input:** Grid descriptor $\mathcal{G}$, rule $f$ with parameters $\theta$, boundary policy $\beta$, device capabilities $\mathcal{C}$, workload profile $\mathcal{W}$, visibility policy $\pi$

**Output:** Versioned state descriptors; timestamp and resource-counter records

VALIDATE($\mathcal{G}, f, \beta$)<br>
$\mathfrak{B} \leftarrow$ SELECTBACKEND($\mathcal{C}, \mathcal{W}$)<br>
$S_A \leftarrow$ INITIALIZESTATE($\mathcal{G}$)<br>
$S_B \leftarrow$ ALLOCATEOUTPUTBUFFER($\mathcal{G}$)<br>
$t \leftarrow 0$<br>
**while** simulation is running **do**<br>
    $\mathcal{Q} \leftarrow$ READVALIDATEDCOMMANDS( )<br>
    APPLYATSTEPBOUNDARY($\mathcal{Q}, \theta$)<br>
    LAUNCHUPDATE($\mathfrak{B}, S_A, S_B, f, \beta$) ▷ $S_B \leftarrow f(S_A; \theta)$<br>
    WAITFORCOMPLETION( ) ▷ compute completion or fence signal<br>
    SWAP($S_A, S_B$); $t \leftarrow t + 1$<br>
    $\delta \leftarrow$ BUILDVERSIONEDDESCRIPTOR($\mathcal{G}, f, t, S_A$)<br>
    PUBLISHASYNC($\delta, S_A$) ▷ non-blocking for the XR frame loop<br>
    **if** RENDERERHASFRESHCOMPLETESTATE( ) **then**<br>
        $S_{\text{vis}} \leftarrow$ ACQUIRELATESTCOMPLETESTATE( )<br>
        UPDATEVISUALIZATION($S_{\text{vis}}, \pi$)<br>
    **end if**<br>
    RECORDTIMESTAMPSANDCOUNTERS( )<br>
**end while**

# EVALUATION

The evaluation has two purposes. First, it assesses whether the proposed architecture addresses the limitations identified in the literature. Second, it defines the measurements needed to test the architecture after implementation. Because no executable implementation or new measurements are supplied with this manuscript, this section reports an analytical evaluation framework, not empirical results. No numerical performance advantage is attributed to the proposed framework at this stage.

## Evaluation dimensions and metrics

Computational throughput. The number of cell updates per second is calculated as

$$R = \frac{N_{cells} \cdot N_{steps}}{T_{run}} \;\; [cell\ updates/s]; \quad R_M = \frac{R}{10^6} \tag{5}$$

where N_cells is the number of updated cells per step, N_steps is the number of simulated steps, and T_run is the measured wall-clock time. The result should be reported as cell updates per second or million cell updates per second (R_M), with the grid size, active-cell ratio, transition rule, and boundary policy stated. Kernel time and end-to-end time should be reported separately.Step-time stability. Report mean, median, 95th percentile, and standard deviation of CA step time. The 95th percentile is important because occasional long updates can create visible stalls even when the mean is acceptable.Rendering performance. Report average FPS, frame-time median, frame-time 95th percentile, missed-frame ratio, and frame-time variance. Average FPS alone is insufficient because it can hide intermittent stutter. XR target rates must be selected according to the actual device and runtime rather than assuming that a single frame-rate threshold is suitable for all headsets.

Latency. At least three latency measures should be distinguished: (i) command-to-commit latency, from user action to application of the command at a simulation boundary; (ii) state-to-visible latency, from completion of a CA step to its presentation in the rendered scene; and (iii) input-to-photon or motion-to-photon latency, where the device and instrumentation permit its measurement. Each measure should specify clock synchronization and timestamp location.

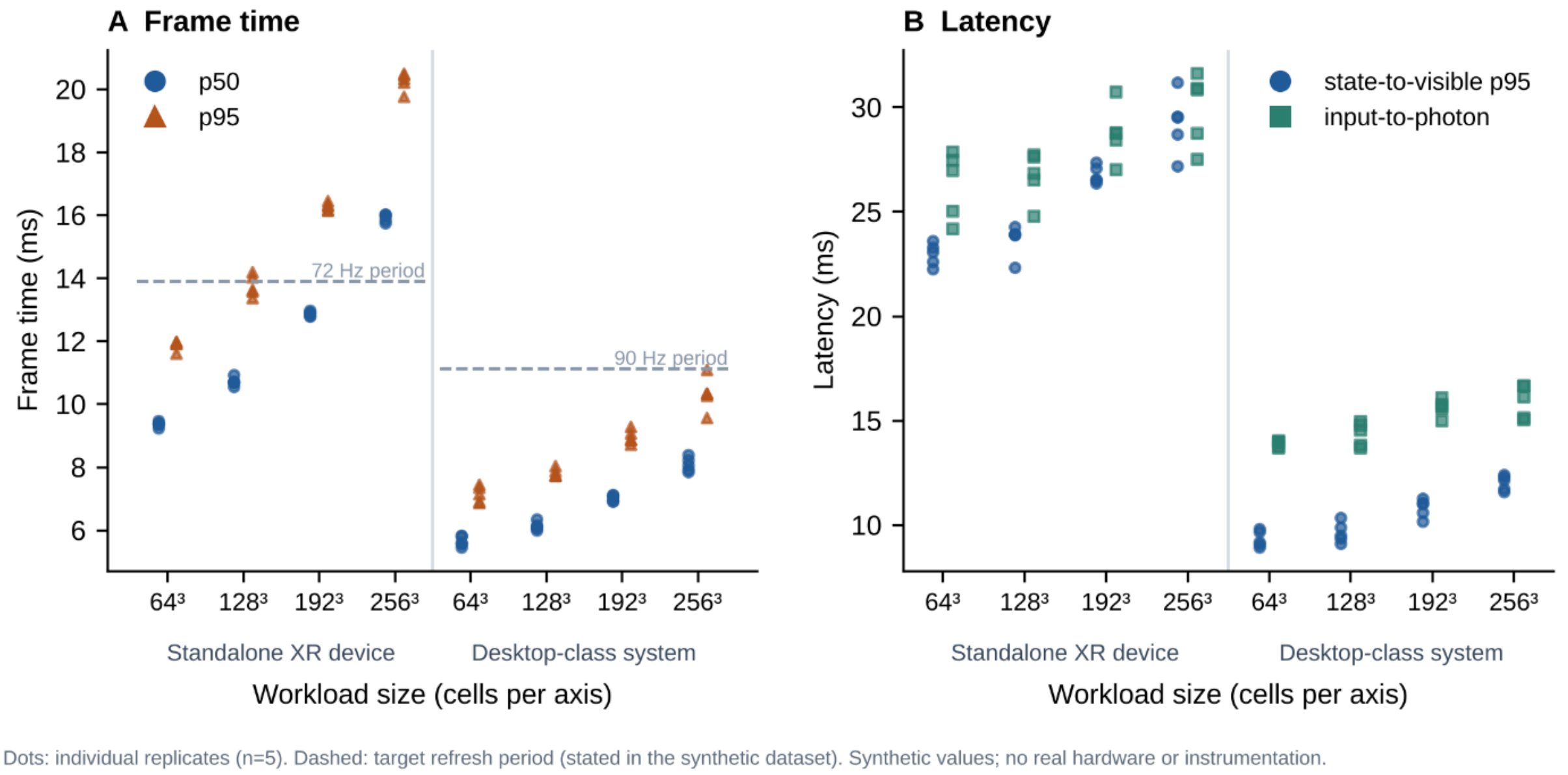

Figure 6. XR frame-time and latency behaviour

Memory use. Record CPU memory, GPU memory, peak allocation, transfer-buffer size, and memory growth over long runs. Dense and sparse representations should be compared under equivalent active-state distributions.

Scalability. Evaluate multiple grid dimensions and active-cell ratios. Strong scaling can be summarized using speedup and parallel efficiency,

$$S(p) = \frac{T_1}{T_p} \tag{6}$$

$$E(p) = \frac{S(p)}{p} = \frac{T_1}{p \cdot T_p} \tag{7}$$

where T_1 is the run time with one processing unit, T_p is the run time with p processing units, and the meaning of p (threads, compute units, or devices) is stated. For heterogeneous platforms, compare end-to-end time under clearly described resource constraints rather than presenting raw device counts as equivalent.

Correctness and reproducibility. Compare optimized backends against the reference implementation using exact state equality for deterministic integer-state CA, or an explicitly defined tolerance for floating-point state. Report divergence over multiple steps, boundary consistency, deterministic repeatability, and preservation of rule semantics.

Interoperability. Measure successful schema negotiation, state-transfer time, serialization overhead, synchronization delay, dropped or stale state ratio, and correctness of coordinate and state interpretation across engines. Compatibility should be demonstrated with actual adapters; a conceptual ability to connect engines is not proof of interoperability.

Interaction quality. Use task-level measures such as task completion time, selection error, parameter-adjustment error, navigation distance, number of corrective actions, and a validated usability questionnaire. User studies should specify participant count, expertise, tasks, training, and desktop comparison conditions. Performance counters and subjective ratings should be reported separately.

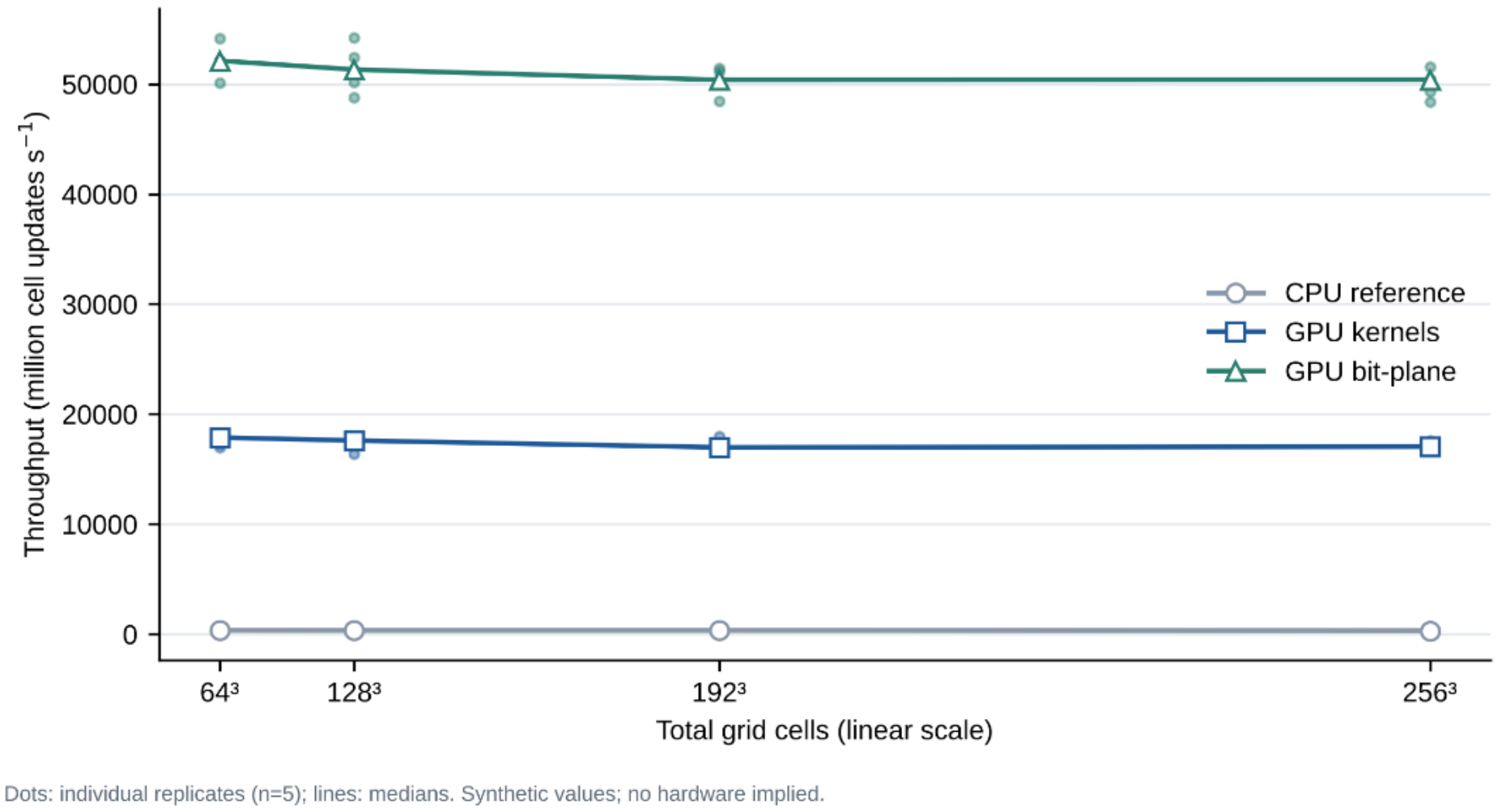

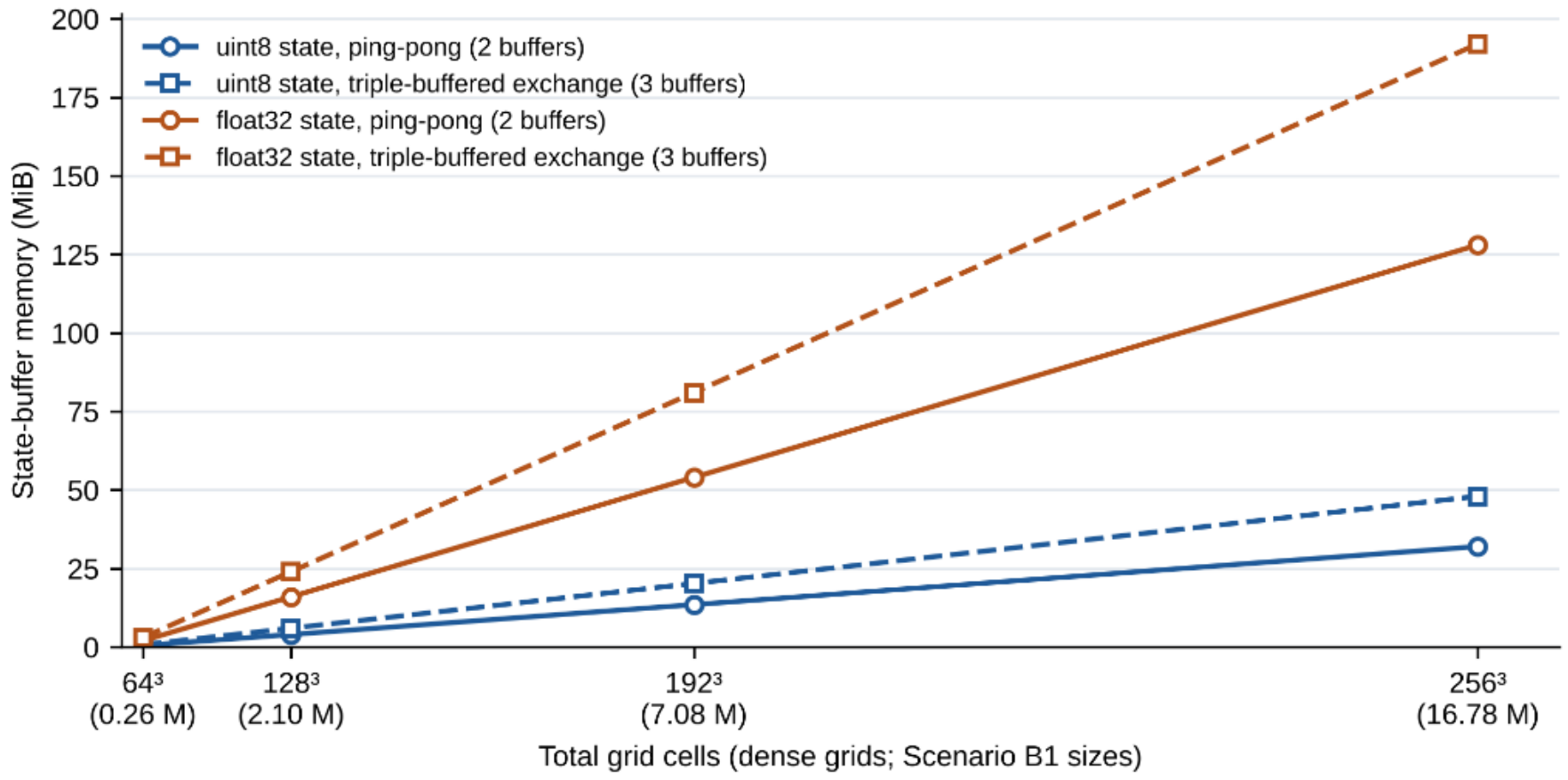

Figure 7,8 – Memory Footprint and Scalability

## Proposed benchmark scenarios

The following scenarios form a proposed benchmark suite (Table 2). They are experimental configurations, not previously measured results.

Table 2. Proposed benchmark suite.

| Scenario | Grid / workload | Main stressor | Required measurements |
|---|---|---|---|
| B1: Dense baseline | Dense cubic grids at $64^3$, $128^3$, $192^3$, and $256^3$ cells | Growth in total cell count | Throughput, step time, CPU/GPU visible latency |
| B2: Sparse activity | Fixed grid with active-cell ratios of 10%, 25%, 50%, and 100% | Sparse-versus-dense representation | Throughput, memory, chunk scheduling overhead |
| B3: Neighbourhood scaling | Same grid with at least three neighbourhood radii or neighbourhood sizes | Increasing neighbourhood work | Step time, throughput, backend suitability |
| B4: XR rendering load | Identical simulation with several visual encodings and view distances | Rendering and fill-rate pressure | FPS, frame-time percentiles, GPU utilization |
| B5: Middleware transfer | Local shared-resource path and serialized-copy path | Data exchange overhead | Transfer time, state freshness, memory copies |
| B6: Long run | Fixed workload | Thermal drift | FPS, step time |
| B7: Interaction tasks | Region selection, slicing, parameter editing, pause/step, annotation | Human interaction and navigation | Completion time, errors, corrective actions |

For a fair comparison, each benchmark should use identical initial states, rule parameters, number of simulation steps, and boundary conditions across implementations. Each configuration should be repeated after a warm-up period. Report median and percentile statistics, not only the best observed run. A second round should repeat selected scenarios on a standalone XR device and a desktop-class system to expose the effects of different power and memory budgets.

## Analytical comparison of architectural options

Table 3 summarizes the expected strengths and risks of the main design options. The table is an analytical comparison derived from the reviewed themes; it is not a measured ranking of implementations.

Table 3. Analytical comparison of architectural options.

| Design option | Potential strength | Principal risk or limitation | Evidence to collect |
|---|---|---|---|
| CPU reference execution | Transparent baseline and easier debugging | Limited parallel throughput for large grids | Reference correctness and step time |
| Conventional GPU kernels | High data parallelism for independent cell updates | Memory bandwidth, transfer, and device dependence | Kernel and end-to-end throughput |
| Bit-plane / bitwise encoding | Efficient state packing and parallel bit operations for compatible rules | Rule-specific complexity and conversion overhead | State equivalence, memory use, speedup |
| Tensor-core execution | Potential acceleration for suitable weighted-neighbourhood formulations | Not applicable to every transition rule; precision and mapping constraints | Rule-specific accuracy and performance |
| Sparse or chunked voxel storage | Avoids processing inactive regions and can support large spatial domains | Indexing, chunk boundaries, and irregular access overhead | Active-ratio scaling and memory footprint |
| Middleware-separated XR | Looser coupling between solver and visualization engine | Synchronization and transport overhead | Cross-engine state consistency and latency |
| Integrated single-engine implementation | Simpler deployment and potentially fewer data-transfer stages | Engine lock-in and reduced solver portability | Porting effort and portability tests |

The table suggests that no single execution or integration strategy is universally optimal. The proposed framework therefore emphasizes an explicit abstraction boundary and workload-aware selection rather than prescribing one acceleration technique for all CA models.

Ablation tests should isolate the contribution of each architectural mechanism. The proposed tests are:

1. A1: Disable asynchronous state exchange and compare frame-time stability and state-to-visible latency.
2. A2: Disable region-of-interest rendering and compare rendering cost at increasing volume sizes.
3. A3: Replace the optimized backend with the CPU reference and quantify the compute contribution separately from middleware overhead.
4. A4: Replace shared-resource exchange with serialization and copying to measure transfer cost.
5. A5: Disable chunking or sparse processing and compare memory pressure and throughput across active-cell ratios.
6. A6: Remove command batching and compare interaction responsiveness and simulation determinism.

An ablation should change one factor at a time while holding the remaining conditions constant. Interactions between factors can be studied in a later factorial design.

## Evaluation outcome expected from the proposed design

(Originally Section 7.8.) The analytical evaluation supports a set of testable expectations, not confirmed results. First, GPU acceleration is expected to reduce CA step time for workloads with sufficient parallelism, but total XR responsiveness will depend on transfer and rendering costs. Second, explicit middleware is expected to improve portability and maintainability, while adding measurable synchronization and data-exchange overhead. Third, asynchronous state publication is expected to reduce blocking in the XR frame loop, but it may increase state age under sustained simulation overload. Fourth, visibility-aware rendering is expected to lower rendering cost for large volumes, but it must not be mistaken for a change in simulation fidelity. Fifth, a rule-independent architecture is expected to simplify domain adaptation, although rule-specific optimized kernels will still require individual validation. These expectations define the hypotheses that the implementation and benchmark study should test.

**Discussion**

The proposed architecture responds to a recurring fragmentation in the reviewed literature. GPU acceleration, volumetric simulation, immersive visualization, and cross-platform XR have each advanced, but their integration introduces costs and correctness requirements that cannot be inferred from the performance of any one component. The framework therefore places equal emphasis on execution and coordination. A high-throughput CA kernel is useful only if its outputs can be transferred, interpreted, and displayed without unacceptable delay; an XR interface is useful only if the visual state remains traceable to a well-defined simulation step.

A central design implication is that hardware abstraction must be workload-aware rather than purely syntactic. A single API can conceal backend differences, but it cannot make all devices computationally equivalent. The framework should expose device capabilities and select among valid implementations while retaining a reference path. Specialized representations such as bit planes or tensor-core computation can be valuable for compatible rules, but they should remain optional backends rather than mandatory assumptions [4, 21]. This allows the architecture to benefit from new hardware without excluding devices that lack a particular feature.

The middleware layer is equally important. Interoperability should mean more than the ability to open the same file or render a similar scene in two engines. It requires preservation of grid dimensions, coordinate systems, time-step identity, transition-rule metadata, and state consistency. MINT demonstrates the value of loosely coupling visualization frameworks and VR environments, while WebGPU-based engines and WebXR interaction toolkits provide additional cross-platform opportunities [10, 17, 31]. The proposed contribution is to make these concerns explicit for CA state exchange.

Another important implication concerns the difference between visual LOD and simulation LOD. A renderer can reduce visual complexity without changing the underlying cellular state. By contrast, reducing the number of simulated cells, merging regions, or changing update frequency can alter the automaton's dynamics. These strategies should be classified separately and subjected to different validation procedures. This distinction is especially important for scientific applications, where a visually plausible output is not necessarily a scientifically equivalent result.

Finally, the evaluation framework treats human interaction as a first-class dimension. Immersive systems should not be justified by frame rate or visual appeal alone. Task-based assessment is needed to determine whether spatial manipulation and navigation improve the user's ability to inspect, understand, and modify complex state spaces. The comparison with desktop workflows should therefore use equivalent tasks and objective outcomes rather than relying only on preference ratings. Existing work on VR exploration of 3D CA and large scientific datasets provides a foundation for such studies, but the interaction hypotheses of the proposed architecture remain to be tested [3, 25, 29].

**Limitations**

This paper has several limitations. First, it is based on a curated reference set supplied for the study rather than a fully documented database search. The search sources, exact search strings, screening records, exclusion decisions, and independent quality assessment are not available; consequently, the review should not be interpreted as exhaustive or PRISMA-compliant.Second, the proposed framework is conceptual. No complete software implementation, XR deployment, or new performance measurement is reported. The architecture's expected benefits—including improved portability, lower synchronization overhead, and more stable frame times—must be verified empirically. The analytical evaluation defines the required metrics but cannot substitute for experimental results.

Third, the reviewed studies use different CA rules, neighbourhoods, hardware, programming models, grid sizes, and measurement procedures. Their performance values cannot be pooled into a single comparative ranking without normalization and access to compatible experimental details. A reported speedup in one workload should not be generalized to every three-dimensional CA model.Fourth, the architecture does not prescribe a single XR engine, graphics API, or compute backend. This improves conceptual portability but leaves implementation choices open. Actual compatibility will depend on the device runtime, graphics and compute APIs, memory-sharing support, and application constraints.

Fifth, the user-interaction model is not validated through a controlled user study. Claims about cognitive offloading, usability, learning, or scientific insight must therefore remain hypotheses until tested with appropriate participants and tasks.

Finally, multi-user synchronization, learned transition rules, and domain-specific scientific validation are treated as extensions rather than fully specified subsystems. Each requires additional design and evaluation. These limitations are consistent with the identified needs for hardware-agnostic scaling, standardized metrics, improved navigation, collaborative synchronization, and rule adaptation.

Future work should proceed along several connected directions.

Hardware-agnostic execution. Implement a backend abstraction that selects CPU, GPU, or specialized kernels according to device capabilities and workload structure. The selection policy should be based on measured end-to-end performance, memory pressure, and energy or thermal constraints rather than kernel speed alone. Standalone mobile XR devices should be included as first-class targets.Standardized CA-XR benchmark suite. Publish a suite containing reference rules, initial conditions, grid configurations, active-cell distributions, correctness tests, and measurement scripts. The suite should record throughput, frame-time distributions, latency, memory, portability, and interaction metrics. This would make it possible to compare frameworks under reproducible conditions.

Cross-engine middleware. Implement at least two independent XR adapters around the same CA state contract. A useful next step is to connect a scientific solver to both a native XR engine and a WebXR/WebGPU client, then measure transfer overhead and state consistency. The contract should be versioned and should include coordinate and time semantics.Adaptive fidelity. Develop region-of-interest policies that allocate rendering detail according to visibility and interaction. A separate research track should investigate approximate simulation LOD while measuring its effect on CA dynamics and domain-level outcomes. Both forms of LOD should be clearly distinguished.

Interaction and navigation. Conduct controlled studies comparing 6-DOF manipulation, slice-based inspection, controller-based selection, and desktop navigation. Tasks should include identifying patterns, comparing time steps, locating propagation fronts, modifying parameters, and tracing the consequences of interventions.Collaborative simulation. Extend the middleware to support multiple clients with a defined authoritative state model, conflict-resolution rules, and recovery behaviour. Evaluate synchronization delay, consistency, and communication overhead as the number of users increases.

Domain-agnostic and data-driven rules. Develop a modular rule registry and investigate learned or hybrid transition functions. Rule adaptation should not require rewriting the visualization or transport layers. Learned rules should be evaluated for reproducibility, interpretability, and generalization in addition to predictive quality.

Longitudinal scientific and educational evaluation. Once a stable implementation is available, assess whether immersive interaction improves task performance, conceptual understanding, or retention compared with desktop alternatives. Such studies should use pre-defined outcomes and suitable follow-up periods rather than relying solely on immediate user satisfaction.

## Conclusion

This paper has presented a hybrid synthesis and a proposed unified architecture for real-time three-dimensional cellular automata simulation in extended reality. The critical review indicates that GPU acceleration, voxel-oriented simulation, scientific visualization in VR, and cross-platform XR each provide important foundations. However, these advances do not automatically form a coherent CA-XR system. The main challenge is to coordinate computation, state exchange, rendering, and interaction while preserving the meaning and reproducibility of the underlying automaton.

The proposed five-layer architecture addresses this challenge by separating voxel and rule management, parallel CA computation, interoperability middleware, visualization and rendering, and XR interaction. Its principal design choices are an explicit state contract, hardware-aware backend selection, asynchronous simulation-rendering decoupling, visibility-aware visual representation, and timestamped interaction commands. The paper further defines a benchmark protocol that measures computational throughput, frame-time stability, latency, memory consumption, scalability, interoperability, correctness, and interaction quality.

The contribution is a structured research and implementation blueprint rather than a claim of completed experimental validation. The next step is to implement the reference engine and at least one optimized backend, connect them to more than one XR client, and conduct the proposed benchmark and user studies. By making both the architecture and its evaluation requirements explicit, the framework provides a basis for developing portable, measurable, and user-centred 3D CA simulation systems for immersive scientific exploration.

## References (APA 7th Edition)

[1] 3d visualization of forest fire spread model based on cellular automata. https://doi.org/10.1109/icccs57501.2023.10150987

[2] Ahmed (n.d.). A framework for interoperability of virtual reality and simulation models in support of training. https://doi.org/10.22215/etd/2012-09681

[3] Arevalo, C., Kariyado, Y., & Villegas, J. (2022). Virtual reality tool for exploration of three-dimensional cellular automata. *Electronics, 11*(3), 497-497. https://doi.org/10.3390/electronics11030497

[4] Brabec, M., Klepl, J., & Kruliš, M. (2026). Improving cellular automata performance with bit-planes encoding and bitwise vectorization. *Parallel Computing, 130,* 103226-103226. https://doi.org/10.1016/j.parco.2026.103226

[5] Czapczyński, M., & Dziedzic, K. (2026). Analysis of latency reduction and performance improvement methods in selected vr applications. *Journal of Computer Sciences Institute, 38,* 87-94. https://doi.org/10.35784/jcsi.8504

[6] Davison, T., Samavati, F., & Jacob, C. (2018). Lifebrush: Painting interactive agent-based simulations. https://doi.org/10.1109/CW.2018.00017

[7] Faldor, M., & Cully, A. (2024). Cax: Cellular automata accelerated in jax. *arXiv.org, abs/2410.02651*. https://doi.org/10.48550/arxiv.2410.02651

[8] Ganji, S. (2026). A unified and practical architecture for xr development. https://doi.org/10.11575/prism/51614

[9] Gao, Y., Yang, J., Xu, X., & Shi, F. (2011). Efficient cellular automaton segmentation supervised by pyramid on medical volumetric data and real time implementation with graphics processing unit. *Expert Systems With Applications, 38*(6), 6866-6871. https://doi.org/10.1016/J.ESWA.2010.12.049

[10] Geringer, S., Geiselhart, F., Bäuerle, A., Dec, D., Odenthal, O. G., Reina, G., Ropinski, T., & Weiskopf, D. (2024). Mint: Integrating scientific visualizations into virtual reality. *Journal of Vision*. https://doi.org/10.1007/s12650-024-01011-y

[11] Gobron, S., Bonafos, H., & Mestre, D. (2008). Gpu accelerated computation and visualization of hexagonal cellular automata. https://doi.org/10.1007/978-3-540-79992-4_67

[12] Gobron, S., Çöltekin, A., Bonafos, H., & Thalmann, D. (2011). Gpgpu computation and visualization of three-dimensional cellular automata. *The Visual Computer, 27*(1), 67-81. https://doi.org/10.1007/S00371-010-0515-1

[13] Hadjar, H., Hadjadj, Z., & Meziane, A. (2026). Human-centered digital twins: A bidirectional webxr-based architecture for immersive interaction. https://doi.org/10.1145/3822516.3834871

[14] Harris, M. J., Coombe, G., Scheuermann, T., & Lastra, A. (2002). Physically-based visual simulation on graphics hardware. https://doi.org/10.5555/569046.569061

[15] Harris, M. J., Coombe, G., Scheuermann, T., & Lastra, A. (2005). Physically-based visual simulation on graphics hardware. https://doi.org/10.1145/1198555.1198791

[16] Heintz, C., Grunwald, M., Edenhofer, S., Hähner, J., & Mammen, S. V. (2017). The game of flow - cellular automaton-based fluid simulation for realtime interaction. https://doi.org/10.1145/3139131.3143415

[17] Hentschel, T., Neuhöfer, J. A., Dresel, M., Klasen, O. B., Goldbach, P., & Jochems, N. (2026). Webxrit - a lightweight interaction toolkit to create virtual reality training applications for webxr with unity eics015. *Proceedings of the ACM on Human-Computer Interaction*, *10*(4), 1-22. https://doi.org/10.1145/3816767

[18] Herruzo, A., Patel, J., Sangaraju, H., Salapu, S. P. K., & Marti, A. O. (2026). Sentient xr: Zero-install browser-based urban digital twins with live environmental sensing. https://doi.org/10.1109/icir70631.2026.11628371

[19] Hnat, B., & Chapman, S. C. (1999). Visualization of multi-scale data sets in self-organized criticality sandpile model. https://doi.org/10.1007/978-3-642-57313-2_23

[20] Kushwah, P. A., Rajguru, T., Bansod, A., Devdhe, P., & Kotala, S. (2026). An immersive ar/vr platform for digital electronics and computer graphics education. https://doi.org/10.5281/zenodo.20771606

[21] Navarro, C. A., Quezada, F. A., Meneses, E., Ferrada, H., & Hitschfeld, N. (2024). Cat: Cellular automata on tensor cores. https://doi.org/10.48550/arxiv.2406.17284

[22] Ohno, N., & KAGEYAMA, A. (2024). Voir: Virtual reality visualization software for large-scale simulations. *Plasma and Fusion Research, 19*, 1401024-1401024. https://doi.org/10.1585/pfr.19.1401024

[23] Olivier, V. A., & Oded, L. (2022). Cellular automata simulation on fpga for training neural networks with virtual world imagery. https://doi.org/10.48550/arxiv.1711.07951

[24] Opencal++: An object-oriented architecture for transparent parallel execution of cellular automata models. https://doi.org/10.1109/pdp59025.2023.00045

[25] Ortiz, J. A., Insley, J. A., Knowles, J. Y. K., Mateevitsi, V., Papka, M. E., & Rizzi, S. (2022). Massive data visualization techniques for use in virtual reality devices. https://doi.org/10.1109/ldav57265.2022.9966400

[26] Pajouheshgar, E., Xu, Y., Mordvintsev, A., Niklasson, E., Zhang, T., & Süsstrunk, S. (2023). Mesh neural cellular automata. *arXiv.org, abs/2311.02820*. https://doi.org/10.48550/arxiv.2311.02820

[27] Pajouheshgar, E., Xu, Y., Mordvintsev, A., Niklasson, E., Zhang, T., & Süsstrunk, S. (2024). Mesh neural cellular automata. *ACM Transactions on Graphics*, *43*(4), 1-16. https://doi.org/10.1145/3658127

[28] Pandit, T., & Kudithipudi, D. (2026). Mini cellular automata for spatio-temporal task identification. https://doi.org/10.1145/3795101.3805424

[29] Pandolfi, R., Todd, J., Donatelli, J. J., & Ushizima, D. (2025). Ascribe-xr: Extended reality for visualization of scientific images. https://doi.org/10.1145/3731599.3767368

[30] Park, S., & Choi, Y. H. (2025). Three-dimensional visualization in biological and medical research: A comparative analysis of blender, unity and unreal engine. https://doi.org/10.22186/jyi.28.12.1.1

[31] Rodríguez, A. M. R., García, P., Marquerie, J. S., Marques, R., & Blat, J. (2025). A cross-platform, webgpu-based 3d engine for real-time rendering and xr applications. https://doi.org/10.1145/3746237.3746305

[32] Shahir, Y. R., & Beigi, A. (2026). Development of 3d neural cellular automata for learning spatio-temporal patterns. https://doi.org/10.1109/iicai70155.2026.11620913

[33] Shen, L., Plessas-Azurduy, P., & Bub, G. (2026). Imyocyte: A web-based cellular automata platform for real-time demonstration of cardiac reentry. *The Biophysicist*. https://doi.org/10.35459/tbp.2026.000295

[34] Sophie, N. (2023). On visualization of three-dimensional cellular automata and tilings with head-mounted display. https://doi.org/10.1007/978-981-99-0688-8_2

[35] Sung, N., Ma, J., Hor, K., Kim, T., Va, H., Choi, Y., & Hong, M. (2025). Real-time physics simulation method for xr application. *Computers, 14*(1), 17-17. https://doi.org/10.3390/computers14010017

[36] Svyetlichnyy, D. S., & Banaś, K. (2026). Conflict-free parallel frontal cellular automata with linked lists for large-scale microstructure simulations. *Journal of Computational Science*, 103024-103024. https://doi.org/10.1016/j.jocs.2026.103024

[37] Trapeznikova, M. A., Yashina, M. V., & Garibyan, A. G. (2024). Software for visualization of cellular automata models in traffic flow analysis. https://doi.org/10.1109/tirved63561.2024.10769828

[38] Waseem, M., & Hong, M. (2025). Moving towards large-scale particle based fluid simulation in unity 3d. *Applied Sciences, 15*(17), 9706-9706. https://doi.org/10.3390/app15179706

[39] Wonnacott, M., Castro, L. E. T., Stroud, H. J., Aisa, L., Dhouioui, M., Williams, R. R., Protopopov, D., Sobrado, S., & Glowacki, D. R. (2026). Nanover: An open-source framework for interactive molecular dynamics in extended reality (imd-xr) on commodity hardware. https://doi.org/10.48550/arxiv.2606.30678

[40] Xu, H., Zlatanova, S., Gorte, B. G. H., Lamsal, R., Heslop, D., Liang, R., & Canbulat, I. (2026). A pilot study on a gpu-accelerated voxel simulation framework for 3d indoor and urban-scale gas dispersion and aerosol transport. *ISPRS International Journal of Geo-Information, 15*(9), 405-405. https://doi.org/10.3390/ijgi15090405

[41] Xu, H., Zlatanova, S., Liang, R., & Canbulat, I. (2026). Heat-transfer-driven voxel-based simulation: An exploratory gpu-accelerated framework for urban-scale 3d fire spread. *ISPRS International Journal of Geo-Information*, *15*(9), 423-423. https://doi.org/10.3390/ijgi15090423

[42] Yu, H., & Chang, T. (2023). Gpu parallelization of particulate matter concentration modeling in indoor environment with cellular automata framework. *Building and Environment*, *243*, 110724-110724. https://doi.org/10.1016/j.buildenv.2023.110724

[43] Zeng, H., Wang, H., & Zhang, B. (2024). A high-performance cellular automata model for urban expansion simulation based on convolution and graphic processing unit. *Transactions in GIS*. https://doi.org/10.1111/tgis.13163

[44] Zhang, J. (2026). Algorithmic parallelization of cellular automata. https://doi.org/10.5281/zenodo.22263230

[45] Zhang, J. (2026). Dynamically adaptive computational geometry using cellular automata. https://doi.org/10.5281/zenodo.22202432

[46] Zhang, J. (2026). Graph-based simulation of cellular automata with dynamic rules. https://doi.org/10.5281/zenodo.22629511

[47] Zhang, J. (2026). Scalable distributed simulation using cellular automata and consensus algorithms. https://doi.org/10.5281/zenodo.22341127

[48] Zhang, Y. (2026). Beyond field lines: High-density 3d magnetic field superposition and multi-dimensional interactive visualization via hardware instancing. https://doi.org/10.1109/icvr70658.2026.11672389

[49] Zhou, X., Rafiee, A., & Oosterom, P. J. M. V. (2026). A high-performance game engine-gis-cfd system for interactive urban wind decision support. *AGILE: GIScience Series, 7*, 1-8. https://doi.org/10.5194/agile-giss-7-49-2026

[50] Savargiv, M., & Bastanfard, A. (2013). Text material design for fuzzy emotional speech corpus based on Persian semantic and structure. 2013 International Conference on Fuzzy Theory and Its Applications (iFUZZY), 380–384. https://doi.org/10.1109/iFuzzy.2013.6825469

[51] Savargiv, M., & Bastanfard, A. (2016). Real-time speech emotion recognition by minimum number of features. 2016 Artificial Intelligence and Robotics (IRANOPEN), 72–76. https://doi.org/10.1109/RIOS.2016.7529493

[52] S.A.H. Minoofam, M.M. Dehshibi, A. Bastanfard, and P. Eftikhari, Ad-hoc Ma'qeli Script Generation Using Block Cellular Automata. J Cell Autom. 7 (2012), pp. 321–334.

[53] Koohestani, B., Rezaei, A., & Bastanfard, A. (2026). Precision through play: Enhancing fundamental laparoscopic skill acquisition via gamified XR and pseudo-haptics. arXiv. https://doi.org/10.48550/arXiv.2611.01229

[54] Rezaei, A., Bastanfard, A., Amirkhani, D., Telvan, N., & Koohestani, B. (2026). Enhancing visual realism in augmented reality models through an interactive sunlight-based lighting and shadowing system. SSRN Electronic Journal. https://doi.org/10.2139/ssrn.6561739